\pdfoutput=1
\documentclass{article}
\usepackage{stywhispers,amsmath,epsfig}
\usepackage[utf8]{inputenc}
\usepackage{booktabs}

\usepackage[expansion=alltext,protrusion=true]{microtype}

\usepackage{xcolor}
\usepackage{graphicx}

\usepackage{array}

\usepackage[algo2e,titlenotnumbered, algoruled]{algorithm2e}

\newcommand{\var}[1]{\mathit{#1}}
\newcommand{\mk}{Mt.\ Kilimanjaro}

\renewcommand\footnotemark{}

\newcolumntype{M}[1]{>{\raggedleft\arraybackslash}p{#1}}

\title{Predicting with limited data -- Increasing the accuracy in VIS-NIR diffuse reflectance spectroscopy by SMOTE}
\twoauthors
  {Christina Bogner}
	{Ecological Modelling\\
	BayCEER, University of Bayreuth\\
	Dr.-Hans-Frisch-Str.~1--3\\
	95445 Bayreuth, Germany}
  {Anna Kühnel, Bernd Huwe}
	{Soil Physics Group\\
	BayCEER, University of Bayreuth\\
	Universitätstr.~30\\
	95447 Bayreuth, Germany}

\begin{document}
%
\maketitle
\begin{abstract}
Diffuse reflectance spectroscopy is a powerful technique to predict soil properties. It can be used \textit{in situ} to provide data inexpensively and rapidly compared to the standard laboratory measurements. Because most spectral data bases contain air-dried samples scanned in the laboratory, field spectra acquired \textit{in situ} are either absent or rare in calibration data sets. However, when models are calibrated on air-dried spectra, prediction using field spectra are often inaccurate. We propose a framework to calibrate partial least squares models when field spectra are rare using synthetic minority oversampling technique (SMOTE). We calibrated a model to predict soil organic carbon content using air-dried spectra spiked with synthetic field spectra. The root mean-squared error of prediction decreased from 6.18 to 2.12 mg g$^{-1}$ and $R^2$ increased from $-$0.53 to 0.82 compared to the model calibrated on air-dried spectra only.

\end{abstract}

\makeatletter{\renewcommand*{\@makefnmark}{}
\footnotetext{\textbf{This is the final version of the full peer-reviewed paper presented at the 6th  GRSS Workshop on Hyperspectral Image and Signal Processing: Evolution in Remote Sensing
(WHISPERS) in Lausanne, Switzerland, June 25--27, 2014.}}\makeatother}
\section{Introduction}
\label{sec:intro}
Diffuse reflectance spectroscopy in the visible and near-infrared range (VIS-NIR DRS) has proved to be useful to assess various soil properties \cite{Stenberg2010}.
It can be employed to provide more data rapidly and inexpensively compared to classical laboratory analysis. Therefore, DRS is increasingly used for vast soil surveys in agriculture and environmental research \cite{Shepherd2007, Vaagen2006}. Recently, several studies have shown the applicability of VIS-NIR DRS \textit{in situ} as a proximal soil sensing technique \cite{ViscarraRossel2009, Waiser2007}. 

To predict soil properties from soil spectra, a model is calibrated, often using partial least squares (PLS) regression. However, when calibration is based on air-dried spectra collected under laboratory conditions, predictions of soil properties from field spectra tend to be less accurate \cite{ViscarraRossel2009}. Usually, this decrease in accuracy is attributed to varying moisture between air-dried calibration samples and field spectra recorded with a variable moisture content. Different remediation techniques have been proposed, ranging from advanced preprocessing of the spectra \cite{Minasny2011} to ``spiking'' the calibration set with field spectra \cite{ViscarraRossel2009}.

In our study, we adopt a slightly different view on the calibration problem. It does not only apply to the varying moisture conditions between the calibration data set and the field spectra. Indeed, it is also valid if we want to predict soil properties in a range where calibration samples are rare. Mining with rarity or learning from imbalanced data is an ongoing research topic in Machine Learning \cite{Weiss2004}. In imbalanced data sets frequent samples outnumber the rare once. Therefore, a model will be better at predicting the former and might fail for the latter.

Two different approaches exist to take care of the data imbalance: we can either adjust the model or ``balance'' the data. The latter approach has the advantage that we can use the usual modelling framework. Synthetic minority oversampling technique (SMOTE) is one way to balance the data. It was first proposed for classification \cite{Chawla2002} and recently for regression \cite{Torgo2013}. SMOTE oversamples the rare data by generating synthetic points and thus helps to equalize the distribution.

In this study, we propose a strategy to increase the prediction accuracy of soil properties from field spectra when they are rare in calibration. The goal of this study is to build a calibration model to predict soil organic carbon content (SOCC) from field spectra by air-dried samples spiked with synthetic field spectra.

\section{Material and methods}
\label{sec:mat_meth}

\subsection{Data acquisition}
\label{ssec:data}
The studied soil was sampled at the southern slopes of \mk{}, Tanzania (3$^\circ$ 4$^ \prime$ 33$^{\prime\prime}$ S, 37$^\circ$ 21$^ \prime$ 12$^{\prime\prime}$ E) in coffee plantations. Due to favourable soil and climate in this region, extensive coffee plantations constitute a frequent form of land use. We took 31 samples for calibration at 4 different study sites. For validation, we scanned 12 field spectra at a wall of a soil pit and sampled soil material for chemical analysis at the scanned spots. We call these validation field spectra F.

After collection, the calibration samples were dried in an oven at 45$^\circ$C and sieved $<$ 2 mm. Subsequently, they were scanned with an AgriSpec portable spectrophotometer equipped with a Contact Probe (Analytical Spectral Devices, Boulder, Colorado) in the range 350--2500 nm with 1 nm intervals. The same spectrometer was used in the field. The instrument was calibrated with a Spectralon white tile before scanning the soil samples. For the measurement, a thoroughly mixed aliquot of the sample was placed in a small cup and the surface was smoothed with a spatula. Each sample was scanned 30 times and the signal averaged to reduce the noise. In the following, we call this calibration data set L.

SOCC was measured in a CNS-Analyser by high temperature combustion with conductivity detectors.

\subsection{Generating data by synthetic minority oversampling}
\label{ssec:smote}
To generate new data to spike the calibration data set L, we used SMOTE \cite{Chawla2002} and its extension for regression \cite{Torgo2013}. This algorithm consists of generating new synthetic data using existing data and is summarized below. In our case, we generated new spectra and the related SOCC using the field spectra F. The new spectra are created by calculating the difference between a field spectrum and one of its nearest neighbours and adding this difference (weighted by a random number between 0 and 1) to the field spectrum. The SOCC of the synthetic spectrum is then a weighted average between the SOCC of the field spectrum and the used nearest neighbour.

SMOTE has two parameters, namely $N$, the number of points to generate for each existing point (given in percent of the whole data set) and $k$, the number of nearest neighbours. To study the influence of these parameters we generated six different synthetic data sets S1 through S6, varying $N=100,200,300$ and $k=3,5$.

\begin{algorithm2e}[htb]
\SetAlgoRefName{}
\caption{SMOTE}
\label{alg:smote}
\DontPrintSemicolon
\KwIn{$T$ original samples to be SMOTEd \;
	\Indp \Indp Amount of SMOTE $N$\% \;
	Number of nearest
neighbours $k$}

\KwOut{($N/100) \times T$ synthetic samples with their target values (i.e.\ concentrations)}

\If{$N<100$}{Randomize the $T$ original samples: \;
	\Indp $T = (N/100) \times T$ \;
$N=100$}

$\var{orig.s}[i]$: original sample $i, i = 1,\dots,T$ \;
$\var{orig.t}[i]$: target value of original sample $i$ \;
$\var{new.s}[j]$: synthetic sample $j, j=1,\dots, (N/100) \times T$ \;
$\var{new.t}[j]$: target values of synthetic sample $j$ \;
$\var{ng}$ $\leftarrow N/100$: number of synthetic samples to compute for each original sample \;

\Indm Generate synthetic samples: \;
\Indp \For{$i$ in $1$ to $T$}%
{$\var{nns} \leftarrow$ compute $k$ nearest neighbours for $\var{orig.s}[i]$ \;
	\For{$\ell$ in $1$ to $ng$}%
	{randomly choose $x \in \var{nns}$ \;
	$\var{diff} = x - \var{orig.s}[i]$ \;
	$\var{new.s}[(i - 1) \times \var{ng} + \ell] = \var{orig.s}[i] + \mathrm{RANDOM}(0,1) \times \var{diff}$
	$d_1 = \mathrm{DIST}(\var{new.s},\var{orig.s}[i])$ \;
	$d_2 = \mathrm{DIST}(\var{new.s},x)$
	$\var{target} = \frac{d_2 \times \var{orig.t}(\var{orig.s}) + d_1 \times \var{orig.t}(x)}{d_1 + d_2}$}}
\Return{$\var{new.t} \cup \var{new.s}$ }
\end{algorithm2e}

\subsection{Data pretreatment and explorative analysis}
\label{ssec:pretreatment}
We corrected each spectrum (calibration, validation and synthetic) for the offset at 1000 and 1830 nm and kept only parts with a high signal-to-noise ratio (450--2400 nm). Then, we transformed the spectra to absorbance $(\log_{10} (\mathrm{1/reflectance}))$ and smoothed them using the Singular Spectrum Analysis (SSA). SSA is a non-parametric technique to decompose a signal into additive components that can be identified as the signal itself or as noise \cite{Golyandina2013}. Finally, we divided each spectrum by its maximum and calculated the first derivative.

In order to assess similarities between the calibration, validation and synthetic data sets, we calculated the Principal Component Analysis (PCA) of the (uncorrected original) spectra L and F and projected the synthetic data into the space spanned by the principal components.

\subsection{Partial least squares regression}
\label{ssec:plsr}
We calibrated seven different PLS models. For model I we used the data set L, the spectra scanned under laboratory conditions. Model II through VII were calibrated on L spiked with synthetic spectra S1 through S6. To find the best model I through VII, we varied the number of PLS components between 1 and 15. Based on the predictions in the leave-one-out cross-validation (LOOCV) we calculated the corrected Akaike Information Criterion \cite{Sugiura1978} AIC$_\mathrm{c} = n \operatorname{ln}(\mathrm{RMSE^2}) + 2p + \frac{2p (p+1)}{n-p-1}$, where $n$ is the number of calibration samples, $p$ the number of PLS components and $\mathrm{RMSE}$ the root mean-squared error. The latter is defined as $\mathrm{RMSE} = \sqrt{\sum_{i=1}^n (\hat{y}_i - y_i)^2}$, where $\hat{y}_i$ are the predicted and $y_i$ the measured SOCCs. We selected the model with the smallest AIC$_\mathrm{c}$ as the most plausible.

To assess the model quality, we used the $\mathrm{RMSE}$, the mean error $\mathrm{ME} = \frac{1}{n} \sum_{i=1}^n \hat{y}_i - y_i$ and the coefficient of determination $R^2 = 1-\sum_{i=1}^n (y_i- \hat{y}_i)^2/\sum_{i=1}^n (y_i- \bar{y})^2$, where $\bar{y}$ is the mean SOCC. 

\subsection{Monte Carlo simulations}
\label{ssec:MC}
SMOTE has two random components because it selects spectra randomly (with replacement) among the nearest neighbours and weights the difference between spectra by a random number (between 0 and 1). To study the influence of these random steps, we generated 100 different datasets S1 through S6. Each data set was then used to spike the calibration data set L, to build a new PLS model and to predict the data set F.

\section{Results and discussion}
\label{sec:results}
\begin{figure}[!ht]
  \centering
\includegraphics[width=0.7\linewidth]{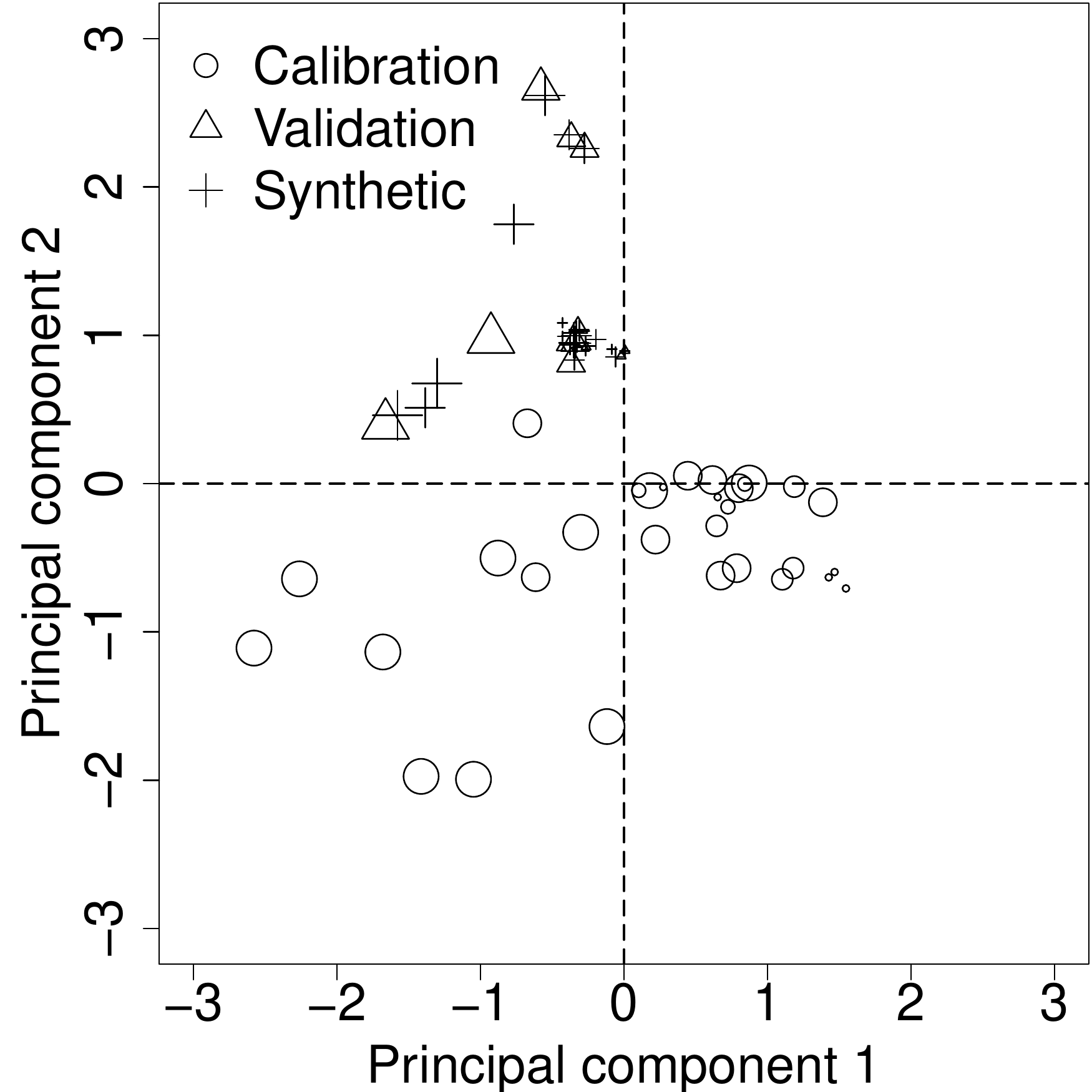}
\caption{Principal component analysis of calibration data set L, validation data set F and one synthetic data set S5. The symbol size was scaled according to the SOCC.}
\label{fig:pca}
\end{figure}

The first principal components (PCs) explain 85.4\% and 11.2\% of variance, respectively. We can clearly identify two distinct groups of samples: the calibration data set L and the field spectra F (Fig.~\ref{fig:pca}). In other words, the data sets L and F differ. The synthetic points that were projected into the space spanned by the PCs resemble the field spectra as expected.

The distinct characteristics of the data sets L and F accord well with the difficulties to predict the data set F by using the laboratory spectra L only (Table~\ref{tab:results:calib} and Table~\ref{tab:results:valid}). Although the LOOCV of model I yields a moderate $\mathrm{RMSE}$ and a large $R^2$, the validation on the data set F fails.

Spiking the calibration data set L with synthetic spectra increases the prediction accuracy of the SOCC in the data set F. Actually, the $\mathrm{RMSE}$ decreases and $R^2$ increases with increasing number of synthetic points both for the LOOCV and the validation (Table~\ref{tab:results:calib} and Table~\ref{tab:results:valid}). However, the number of model parameters also increases from 2 to 7.

The Monte Carlo results show only a small variability in the interquartile range. However, some synthetic data sets in model V produced $R^2$ values smaller than $-$0.53, the value we obtain in model I on air-dried samples only. This might be due to the combination of neighbours during smoting. In general, models with 5 neighbours were more accurate than those with 3 neighbours. However, the number of neighbours had a smaller influence on the prediction accuracy than the number of synthetic points.

It is difficult to decide \textit{a priory} how many synthetic points should be included in the calibration. Indeed, in a classification problem the goal is to approximate an equal distribution of different classes such that the rare class becomes an ordinary one. In regression, however, we do not know which features of the data make them rare. For our data, the range of SOCC in the data set L is larger than in the data set F. Therefore, we conclude that concentration is not responsible for the difference between these data sets.

\begin{figure}[!ht]
  \centering
\includegraphics[width=\linewidth]{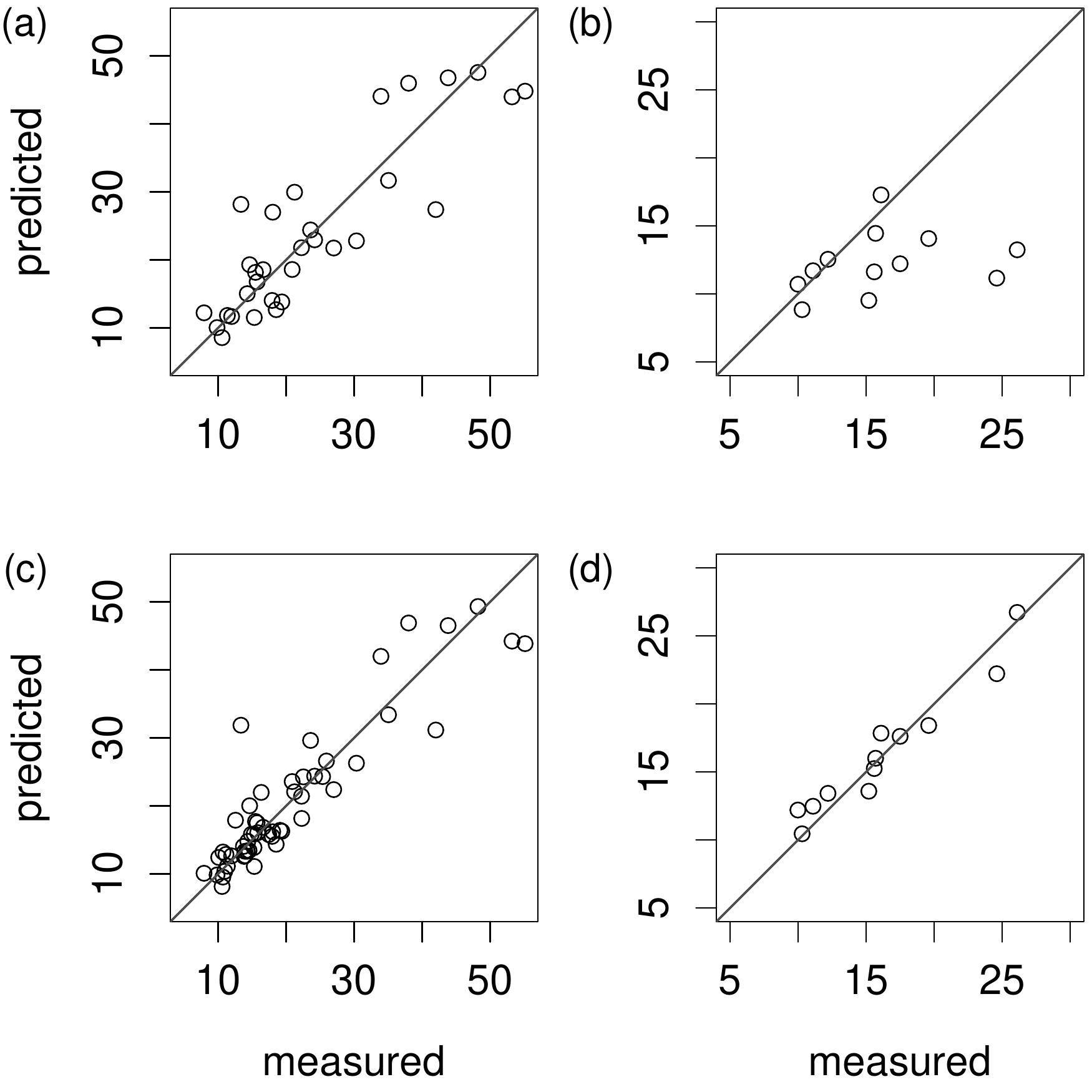}
\caption{Results of (a) leave-one-out cross-validation on data set L (model I), (b) validation on data set F, (c) leave-one-out cross-validation on data set L spiked with a synthetic data set (model VI) and (d) validation on data set F.}
\label{fig:models}
\end{figure}

Based on the Monte Carlo results we chose one synthetic data set from model VI, namely the one with the median number of model parameters and the best $R^2$ in the validation. Thus, the calibration data set includes 31 air-dried and 24 synthetic spectra. Compared to model I, spiking the air-dried data set L with these synthetic spectra clearly improves the prediction of the data set F (Fig.~\ref{fig:models}).

\begin{table*}[htp]
  \centering
\caption{Statistics of the PLS calibration. Median values and 25\% and 75\% quantiles in parenthesis.}
\label{tab:results:calib}
\begin{tabular}{l l r r r r r r }
\toprule
Model & Data set(s) & $N (\%)$ & $k$ & \multicolumn{1}{c}{$p$} & $\mathrm{RMSE}$ (mg g$^{-1}$) & \multicolumn{1}{c}{$R^2$} & \multicolumn{1}{c}{ME (mg g$^{-1}$)} \\ \midrule
I & L & -- & -- & 2\phantom{ (6; 6)} & 6.25\phantom{ (4.45; 4.55)} & 0.77\phantom{ (4.45; 4.55)} & $-$0.20\phantom{ (\phantom{$-$}4.45; \phantom{$-$}4.55)} \\
II & L and S1 & 100 & 3 & 5 (4; 5) & 5.29 (5.18; 5.47) & 0.80 (0.79; 0.81) & $-$0.06 ($-$0.10; $-$0.01) \\
III & L and S2 & 200 & 3 & 6 (6; 6) & 4.51 (4.47; 4.56) & 0.83 (0.83; 0.84) & 0.07 (\hphantom{$-$}0.03; \hphantom{$-$}0.11) \\
IV & L and S3 & 300 & 3 & 7 (6; 7) & 4.01 (3.98; 4.06) & 0.85 (0.84; 0.85) & 0.08 (\hphantom{$-$}0.05; \hphantom{$-$}0.11) \\
V & L and S4 & 100 & 5 & 4 (3; 5) & 5.31 (5.16; 5.55) & 0.80 (0.78; 0.81) & $-$0.02 ($-$0.10; \hphantom{$-$}0.04) \\
VI & L and S5 & 200 & 5 & 6 (6; 6) & 4.51 (4.45; 4.55) & 0.83 (0.83; 0.84) & \hphantom{$-$}0.06 (\hphantom{$-$}0.01; \hphantom{$-$}0.10) \\
VII & L and S6 & 300 & 5 & 6 (6; 7) & 4.05 (4.02; 4.08) & 0.84 (0.84; 0.85) & \hphantom{$-$}0.07 (\hphantom{$-$}0.05; \hphantom{$-$}0.09) \\
\bottomrule

\end{tabular}
\end{table*}

\begin{table*}[htp]
\centering
\caption{Statistics of the PLS validation. Median values and 25\% and 75\% quantiles in parenthesis.}
\label{tab:results:valid}
\begin{tabular}{l r r r }
\toprule
Model & $\mathrm{RMSE}$  (mg g$^{-1}$) & \multicolumn{1}{c}{$R^2$} & \multicolumn{1}{c}{ME  (mg g$^{-1}$)}\\ \midrule
I & 6.18\phantom{ (1.81; 2.39)} & $-$0.53\phantom{ (0.77; 0.87)} & $-$3.88\phantom{ ($-$0.77; 0.87)} \\
II & 3.09 (2.82; 3.58) & 0.62 (0.49; 0.68) & $-$0.03 ($-$0.53; 0.79) \\
III  & 2.00 (1.79; 2.40) & 0.84 (0.77; 0.87) & 0.14 ($-$0.01; 0.36) \\
IV  & 1.31 (1.08; 1.58) & 0.93 (0.90; 0.95) & 0.16 (\phantom{$-$}0.06; 0.27) \\
V  & 3.06 (2.79; 3.56) & 0.62 (0.49; 0.69) & $-$0.28 ($-$0.70; 0.79) \\
VI  & 2.12 (1.81; 2.39) & \phantom{$-$}0.82 (0.77; 0.87) & \phantom{$-$}0.24 ($-$0.04; 0.48) \\
VII  & 1.62 (1.29; 2.07) & \phantom{$-$}0.89 (0.83; 0.93) & \phantom{$-$}0.18 (\phantom{$-$}0.02; 0.37) \\
\bottomrule 

\end{tabular}
\end{table*}

\section{Conclusions}
\label{sec:conclusions}
We propose a framework to predict soil properties from \textit{in situ} acquired field spectra by spiking air-dried laboratory calibration data by synthetic ones generated from these field spectra. In general, the prediction accuracy increases when a sufficient number of synthetic points is included in the calibration. However, because it is difficult to determine this number \textit{a priori}, we recommend to generate several synthetic data sets to find an appropriate model.

\begin{center} \textbf{ACKNOWLEDGEMENTS} \end{center}

This study is part of the project DFG FOR 1246 ``Kilimanjaro ecosystems under global change: Linking biodiversity, biotic interactions and biogeochemical ecosystem processes'' and was supported by the Deutsche Forschungsgemeinschaft.





\begin{thebibliography}{10}

\small 

\bibitem{Stenberg2010}
B.~Stenberg and R.~A. Viscarra~Rossel,
\newblock ``Diffuse reflectance spectroscopy for high-resolution soil
  sensing,''
\newblock in {\em Proximal Soil Sensing}, R.~A. Viscarra~Rossel, A.~B.
  McBratney, and B.~Minasny, Eds., pp. 29--47. Springer, 2010.

\bibitem{Shepherd2007}
K.~D. Shepherd and M.~G. Walsh,
\newblock ``Infrared spectroscopy - enabling an evidence-based diagnostic
  surveillance approach to agricultural and environmental management in
  developing countries,''
\newblock {\em Journal of near Infrared Spectroscopy}, vol. 15, no. 1, pp.
  1--19, 2007.

\bibitem{Vaagen2006}
T.-G. V{\aa}gen, K.~D. Shepherd, and M.~G. Walsh,
\newblock ``{Sensing landscape level change in soil fertility following
  deforestation and conversion in the highlands of Madagascar using Vis-NIR
  spectroscopy},''
\newblock {\em Geoderma}, vol. 133, no. 3, pp. 281--294, 2006.

\bibitem{ViscarraRossel2009}
R.~A. Viscarra~Rossel, S.~R. Cattle, A.~Ortega, and Y.~Fouad,
\newblock ``In situ measurements of soil colour, mineral composition and clay
  content by vis--nir spectroscopy,''
\newblock {\em Geoderma}, vol. 150, no. 3, pp. 253--266, 2009.

\bibitem{Waiser2007}
T.~H. Waiser, C.~L.~S. Morgan, D.~J. Brown, and C.~T. Hallmark,
\newblock ``In situ characterization of soil clay content with visible
  near-infrared diffuse reflectance spectroscopy,''
\newblock {\em Soil Science Society of America Journal}, vol. 71, no. 2, pp.
  389--396, 2007.

\bibitem{Minasny2011}
B.~Minasny, A.~B. McBratney, V.~Bellon-Maurel, J.-M. Roger, A.~Gobrecht,
  L.~Ferrand, and S.~Joalland,
\newblock ``Removing the effect of soil moisture from nir diffuse reflectance
  spectra for the prediction of soil organic carbon,''
\newblock {\em Geoderma}, vol. 167, pp. 118--124, 2011.

\bibitem{Weiss2004}
G.~M. Weiss,
\newblock ``Mining with rarity: A unifying framework,''
\newblock {\em Sigkdd Explorations}, vol. 6, pp. 1--19, 2004.

\bibitem{Chawla2002}
N.~V. Chawla, K.~W. Bowyer, L.~O. Hall, and W.~P. Kegelmeyer,
\newblock ``Smote: Synthetic minority over-sampling technique,''
\newblock {\em Journal of Artificial Intelligence Research}, vol. 16, pp.
  321--357, 2002.

\bibitem{Torgo2013}
Lu{\'\i}s Torgo, Rita~P Ribeiro, Bernhard Pfahringer, and Paula Branco,
\newblock ``Smote for regression,''
\newblock in {\em Progress in Artificial Intelligence}, pp. 378--389. Springer,
  2013.

\bibitem{Golyandina2013}
Nina Golyandina and Anatoly Zhigljavsky,
\newblock {\em Singular spectrum analysis for time series},
\newblock Springer, 2013.

\bibitem{Sugiura1978}
N.~Sugiura,
\newblock ``Further analysts of the data by akaike's information criterion and
  the finite corrections,''
\newblock {\em Communications in Statistics-Theory and Methods}, vol. 7, no. 1,
  pp. 13--26, 1978.

\end{thebibliography}

\end{document}